\documentclass{article}
\usepackage{nac_preprint}
\usepackage{comment}
\usepackage{enumitem}
\usepackage{times}  
\usepackage{helvet}  
\usepackage{courier}  
\usepackage[hyphens]{url}  
\usepackage{amsmath,mathtools}
\usepackage{xcolor}
\usepackage{graphicx} 
\usepackage{caption} 
\usepackage{wrapfig}
\usepackage{algorithm}
\usepackage{algorithmic}

\usepackage{newfloat}
\usepackage{listings}
\DeclareCaptionStyle{ruled}{labelfont=normalfont,labelsep=colon,strut=off} 
\floatstyle{ruled}
\newfloat{listing}{tb}{lst}{}
\floatname{listing}{Listing}
\title{A Neuro-mimetic Realization of the Common \\Model of Cognition via Hebbian Learning \\and Free Energy Minimization}

\author{%
  Alexander Ororbia\thanks{These authors contributed equally.} \\
  Rochester Institute of Technology \\
  Rochester, NY, USA \\
  \texttt{ago@cs.rit.edu}
  \And
  Mary Alexandria Kelly\footnotemark[1] \\
  Carleton University \\
  Ottawa, ON, Canada \\
  \texttt{mary.kelly4@carleton.ca}
}

\begin{document}

\maketitle

\begin{abstract}
Over the last few years, large neural generative models, capable of synthesizing semantically rich passages of text or producing complex images, have recently emerged as a popular representation of what has come to be known as ``generative artificial intelligence'' (generative AI). Beyond opening the door to new opportunities as well as challenges for the domain of statistical machine learning, the rising popularity of generative AI brings with it interesting questions for Cognitive Science, which seeks to discover the nature of the processes that underpin minds and brains as well as to understand how such functionality might be acquired and instantianted in biological (or artificial) substrate. With this goal in mind, we argue that a promising research program lies in the crafting of cognitive architectures, a long-standing tradition of the field, cast fundamentally in terms of neuro-mimetic generative building blocks. Concretely, we discuss the COGnitive Neural GENerative system, such an architecture that casts the Common Model of Cognition in terms of Hebbian adaptation operating in service of optimizing a variational free energy functional.

\end{abstract}

\section{Introduction}
\label{sec:intro}

Thanks to deep learning, the state-of-the-art in artificial intelligence (AI) has been advancing rapidly. Many problems deemed impossible a decade ago can now be performed by AI at an expert-level, such as complex board games and video games including Go \cite{silver2016alphago} and StarCraft \cite{vinyals2019starcraft}. For creative activities like art, writing, and conversation, it has become difficult to distinguish AI from human performance, leading to generative AI winning art contests \cite{roose2022art} and, conversely, human artists being mistaken for AI \cite{stokel-walker2023art}. Between tech CEOs claiming significant strides towards artificial general intelligence (AGI) and Geoffrey Hinton, ``godfather of AI,'' warning of a potential ``risk to humanity'' \cite{metz2023ai}, it can be easy to believe that human-level AI is imminent.

It is the role of the cognitive scientist to ask critical questions of the recent developments in artificial intelligence:
\begin{enumerate}
\item Do humans and current AI agents behave similarly?
\item In what ways do humans and current AI agents differ?
\item Where and why do current AI agents and humans differ?
\item How can the differences be bridged?
\end{enumerate}
Our focus is the fourth question. Where there is a gap between current AI and human behaviour, can we bridge such a gap by exploring novel architectures? We are working towards one possible answer to this question in form of a new cognitive architecture, CogNGen (the COGnitive Neural GENerative system; \cite{ororbia2021towards,ororbia2022cogngen,ororbia2023maze}). CogNGen is built on two neurobiologically and cognitively plausible models, namely a variant of predictive processing \cite{clark2015surfing,salvatori2023brain} 
known as neural generative coding (NGC; \cite{ororbia2020continual}) and vector-symbolic (a.k.a. hyperdimensional computing; \cite{Gayler2003,Kanerva2009}) models of human memory \cite{Hintzman1986,KellyMewhortWest}. Desirably, the use of these particular building blocks yields scalable, local update rules, based on variants of Hebbian learning \cite{hebb1949organization}. Such a form of plasticity allows the system to adapt its parameters while facilitating robustness in acquiring, storing, and composing distributed representations of encountered tasks.

In what follows, we motivate CogNGen by situating it within the Common Model of Cognition and recent generative neural network research. We provide an overview of the CogNGen architecture, discuss preliminary results on maze-learning tasks, and detail future directions for CogNGen's development. By explaining our approach at a high level, we aim to both motivate our own research and provide inspiration for other research programs that seek to progressively work towards bridging the gap between humans and AI. 

\section{The Common Model of Cognition}
\label{sec:cmc}

\begin{figure}
\begin{center}
\includegraphics[width=0.55\linewidth]{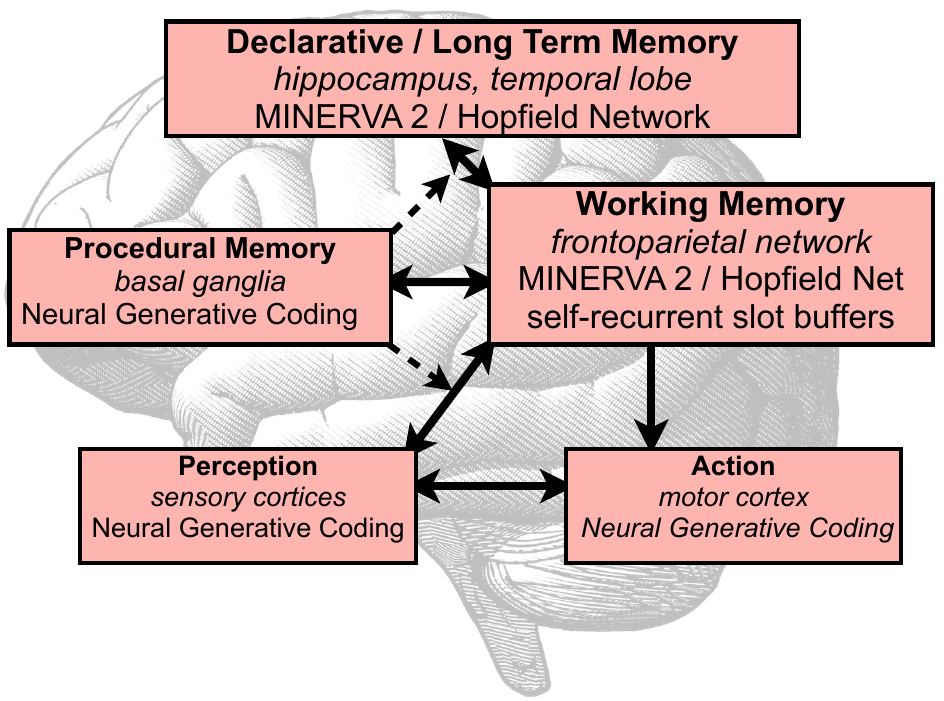}
\end{center}
\caption{Common Model of Cognition \protect\cite{laird2017standard}, associated brain areas \protect\cite{Stocco2021cmc,Stein-Hanson2018,Stocco2018}
and our approach to modelling each module. Solid arrows pass data while dashed arrows modulate data passing.
}
\label{FigCMC}
\vspace{-0.5cm}
\end{figure}

Cognitive architectures serve as unified theories of cognition and as computational frameworks for implementing models of specific tasks. Countless architectures have emerged over the past forty years, often with many  similarities to one another \cite{Kotseruba2018}. On the basis of the strong commonalities between existing cognitive architectures, \cite{laird2017standard} propose the \emph{Common Model of Cognition} (CMC), a high-level theory of the modules of the mind and their interactions (see Fig. \ref{FigCMC}).

The Common Model of Cognition describes several key modules: perceptual and motor modules for interacting with the agent's environment, short-term/working memory buffers for holding the active data in the agent's mind, a declarative (long-term) memory that stores world knowledge, and a procedural memory that controls information flow and evaluates possible actions \cite{laird2017standard}. An evaluation of fMRI data taken from $200$ participants across a diverse of tasks found correlations in patterns of activity across brain areas consistent with the Common Model of Cognition \cite{Stocco2021cmc}.

Although research in cognitive psychology and cognitive neuroscience supports the Common Model of Cognition (CMC, Common Model) as a framework for understanding cognition, is the Common Model the correct framework for developing a novel cognitive architecture that is based on modern machine learning techniques? Our answer is \textit{yes} and, as such, we have modelled CogNGen after the CMC, as illustrated in Figure~\ref{FigCMC}. However, to further motivate our choice of adopting the CMC as a framework, we will next discuss the current state of generative networks and situate the CMC within current machine learning approaches.

\section{Attention is All You Need}
\label{sec:attention}

In \cite{Vaswani2017transformer}, \textit{Attention is All You Need}, a new type of neural model was proposed, the transformer, which demonstrated superior performance on natural language processing (NLP) tasks when compared to recurrent networks, which had been the dominant paradigm for connectionist approaches to NLP since the \cite{elman1990} network. The crucial difference between the transformer and earlier models was the inclusion of a large number of ``attention layers'' to facilitate the selective weighting of different words or sub-word language units based on contextual relevance. 

The claim that attention is \textit{all} that is needed for NLP is a bold one. Yet, chatbots powered by transformers (e.g., OpenAI's ChatGPT, Microsoft's BingChat, Google's Bard) are fluent and versatile conversational partners. OpenAI's GPT-4 is able to write essays and solve math problems at the level of a C+ first-year student \cite{rudolph2023chatbots}. 
Notably, attention has not just been useful for NLP, as the transformer has also achieved state-of-the-art performance in visual \cite{bi2021vision} and audio processing tasks \cite{lin2022transformers,verma2021audio}. The ability of transformers to communicate as if they ``understand'' has even raised questions as to whether transformers could have consciousness \cite{chalmers2023llm}. In \cite{juliani2022perceiver}, Perceiver, a variant transformer model, was evaluated on a variety of working memory and selective attention tasks; it was found that the Perceiver implements Global Workspace Theory \cite{baars1993gwt,mashour2020gwt}, a leading theory of access consciousness.

It should be noted, however, that the Perceiver architecture adds more than just ``attention.'' Perceiver consists of input and ouptut modules, a shared central workspace for data, and attentional mechanisms that select data from the modules, modify the contents of the workspace, as well as pass data from the workspace back to the modules. Thus, the Perceiver has almost all of components of the Common Model of Cognition (see Figure \ref{FigCMC}): perception (input) and motor (output) modules, a working memory (a shared workspace), and the ability to modulate the flow of data in and out of the workspace (procedural memory). Perceiver lacks only the CMC's declarative memory. Crucially, \cite{juliani2022perceiver} found that the removal of any single component causes the Perceiver to fail on some of the working memory and selective attention tasks and thus fails to satisfy the criteria for implementing Global Workspace theory. As a result, all CMC components (except for declarative memory) are \textit{necessary} to implement the functionalities of human consciousness, as postulated by Global Workspace Theory. 

While \cite{juliani2022conscious} found that state-of-the-art networks satisfy many of the criteria for consciousness across different theories, many other capacities associated with consciousness are absent, chiefly episodic memory, which is necessary for the sense of self and history required for self-awareness and long-term planning. It was noted in \cite{juliani2022conscious} that generative networks could be augmented with a memory network to address the lack of a long-term memory. Yet, it was also pointed out in \cite{juliani2022conscious} that other capacities are required to achieve human-like consciousness, including theory of mind, causal reasoning, and meta-cognition. Thus, while adding attention layers to modern neural networks has led to tremendous progress, attention is \textit{not} all that is needed for generative networks to achieve human-level performance. Rather, all components of the CMC are necessary for generative networks to exhibit human-level performance, and further components not specified by the CMC may also be necessary.

\section{What is Attention?}
\label{sec:what_is_attention}

It is undeniable that the addition of attention layers has resulted in a significant step forward in the capabilities of deep learning. 
Even so, \cite{Vaswani2017transformer} did not set out to evaluate the transformer as a model of human attention; rather, the term \textit{attention layer} was a mere act of naming a useful computational mechanism.

Is the ``attention'' in \cite{Vaswani2017transformer} what a cognitive scientist would call attention? Attention is nebulously defined in cognitive psychology. However, a mathematical analysis of the attention layer yields a correspondence to a real mechanism in the brain. In \cite{hebb1949organization}, it was critically observed of neurons that:
\begin{quote}
\small{
When an axon of cell A is near enough to excite a cell B and repeatedly or persistently takes part in firing it, ... A's efficiency, as one of the cells firing B, is increased}
\end{quote}
\noindent Or, as Hebbian learning is pithily summarized, ``neurons that fire together wire together.'' Each occurrence of an input pattern strengthens the ability of the network to reproduce it later. In effect, Hebbian learning allows neural networks to encode memories in their synaptic connectivity.

An early use of Hebbian learning is the \cite{hopfield1982} network. Given a Hopfield network that has memorized \textit{m} patterns (i.e., traces) and a pattern $\mathbf{m_p}$ that is the input to the network (i.e., probe), an \textit{echo} $\mathbf{m_e}$ is retrieved as a sum of all memory traces, each trace weighted by its similarity to the probe as measured by the $\cdot$ dot product. Formally, this is computed as follows:
\begin{equation}
\mathbf{m_e}=\sum_{i=1}^{m}(\mathbf{m_p} \cdot \mathbf{m}_i) \mathbf{m}_i \label{eqn:hopfield_retrieve}
\end{equation}
\noindent where $\mathbf{m}_i$ is the \textit{i}-th trace memorized by the network. 
Unfortunately, the memory storage capacity of the \cite{hopfield1982} network is very limited. As more traces are stored, the more likely the network is to return an echo that is a confused muddle of different traces. The storage capacity can be improved by adding non-linearity -- we can raise the similarities between the probe and traces to an exponent $b$, such that highly similar memory traces contribute disproportionately more to the echo than dissimilar memory traces:
\begin{equation}
\mathbf{m_e}=\sum_{i=1}^{m}(\mathbf{m_p} \cdot \mathbf{m}_i)^b \mathbf{m}_i . \label{eqn:minerva_retrieve}
\end{equation}
\noindent With the addition of the exponent \textit{b} we have the memory retrieval equation for the MINERVA~2 model of human memory \cite{Hintzman1986}. The model in \cite{Hintzman1986} is a Hopfield network with a non-linear weighting of the memories to improve retrieval \cite{KellyMewhortWest}. 
Over the $35$ years of research since it was first proposed, MINERVA~2 has accounted for human memory and learning in experimental paradigms too numerous to exhaustively cite, including judgement of frequency and recognition \cite{Hintzman1984}, 
category learning \cite{Hintzman1986}, implicit learning 
\cite{Jamieson2011}, 
associative learning in humans and animals 
\cite{Jamieson2012}, heuristics and biases in decision-making \cite{Dougherty1999}, hypothesis-generation \cite{Thomas2008}, 
word learning \cite{Jamieson2018semantic}
, and sentence production \cite{johns2016combinatorial,Kelly2020indirect}.

Hopfield networks can achieve exponential storage capacity by using an exponential function of the similarity to weight memory traces \cite{demircigil2017hopfield}. We can adopt the widely-used \textit{softmax} function, or normalized exponential function, for the purpose of weighting memory traces at retrieval \cite{ramsauer2021hopfield}:
\begin{equation}
\mathbf{m_e}=\sum_{i=1}^{m} \frac{(e^{\mathbf{m_p} \cdot \mathbf{m}_i})}{\sum_{j=1}^{m}(e^{\mathbf{m_p} \cdot \mathbf{m}_j})} \mathbf{m}_i . \label{eqn:transformer_retrieve}
\end{equation}

\noindent Modern Hopfield networks use the \textit{softmax} to weight memory traces for efficient retrieval, which is also \textit{exactly} what an attention layer in a transformer uses to apply attentional weighting across a set of items \cite{ramsauer2021hopfield}. As such, transformers can be best understood as networks with a mix of conventional, back-propagation (backprop) trained layers and Hopfield memory layers whose contents are rapidly updated using Hebbian learning. Desirably, upon presentation of input pattern $\mathbf{m_p}$, the energy functional $\mathcal{F}$ associated with a Hopfield network can be formulated as:
\begin{align}
    \mathcal{F}(\Theta) = \frac{1}{2} \sum^{N_f}_{i=1} (\mathbf{m_p})^2_i - \log\bigg( \sum^{m}_{k=1} \exp\Big(\sum^{N_f}_{i=1} \mathbf{M}_{k i} (\mathbf{m_p})_i \Big) \bigg)
\end{align}
where $N_f$ is the number of input features, $\Theta = \{\mathbf{M}\}$, and $\mathbf{M}$ is the memory matrix of currently memorized patterns. 

While backprop is a powerful learning algorithm, as yet, there is no consensus on identifying a neurological equivalent to backprop in the brain \cite{crick1989recent}. The primary challenge to realizing back-propagation of the error signal in the brain is that it requires that non-local information be used to modify synaptic connectivity globally in a highly coordinated manner. A number of potential biological mechanisms have been proposed, such as error signalling via neuropeptides \cite{liu2022neuropeptides} or high-frequency bursts of spikes from pyramidal neurons \cite{payeur2021burst}. Alternatively, one can commit to strictly local synaptic update rules that can approximate the optimization of backprop (e.g., \cite{betti2020backprop}), which is the approach we adopt \cite{ororbia2020continual}. Furthermore, when a network is trained sequentially using backprop  to learn a series of distinct tasks, \textit{catastrophic interference} occurs, with each new task causing the network to forget all previous tasks \cite{french1999catastrophic,mannering2021catastrophic,mccloskey_catastrophic_1989}. 

To avoid the neural and behavioural pitfalls of backprop, we only use Hebbian learning in CogNGen. For working memory and declarative memory, we use MINERVA~2, implemented as Equation~\ref{eqn:minerva_retrieve} with $b=100$ for the simulations discussed in this paper, though future versions of the model will use the more efficient \textit{softmax} retrieval as in Equation~\ref{eqn:transformer_retrieve}. As a complement to MINERVA~2's Hebbian learning, we craft neural circuits that also minimize their own energy functionals derived from the free energy principle \cite{friston2005theory,friston2010free}, instantiated as a form of predictive coding \cite{rao1999predictive,ororbia2020continual,salvatori2023brain}.

\begin{figure}[!t]
\begin{center}
\includegraphics[width=0.425\linewidth]{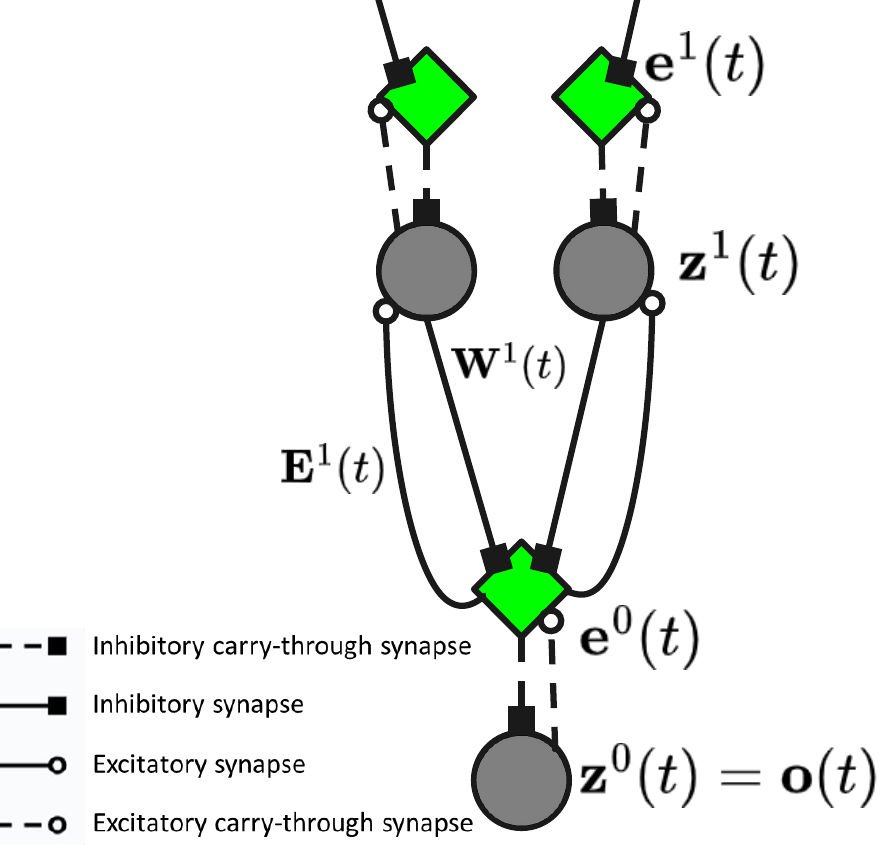}
\end{center}
\caption{NGC circuit with state units $\{\mathbf{z}^0(t),\mathbf{z}^1(t)\}$ (dark grey circles), error neurons $\{\mathbf{e}^0(t),\mathbf{e}^1(t)\}$ (green diamonds), generative synapses $\mathbf{W}^1(t)$, and feedback synapses $\mathbf{E}^1(t)$.
}
\label{fig:ngc}
\end{figure}

\section{Free Energy and Predictive Processing}
\label{sec:fep_pc}

One of the two neural building blocks of CogNGen (alongside Hopfield networks), and the one that also endows it with a fundamental generative quality, is predictive coding (PC; \cite{clark2015surfing,friston2005theory}), based on a generalization known as neural generative coding (NGC; \cite{ororbia2022neural}). NGC circuits can be expressed in terms of the quantity that they optimize at each step in time, an objective function known as the variational free energy (VFE; \cite{friston2010free}). Formally, the VFE functional can be written down as:
\begin{equation}
    E(p,q,\mathbf{o}) = \underbrace{\sum_{\mathbf{z}}q(\mathbf{z}) \text{log}\bigg( \frac{q(\mathbf{z})}{p(\mathbf{z})} \bigg)}_{\text{Complexity Term}} \ + \ \underbrace{\sum_{\mathbf{z}}q(\mathbf{z})\text{log}\bigg(\frac{1}{p(\mathbf{o} \mid \mathbf{z})}\bigg)}_{\text{Accuracy Term}}  \label{eqn:free_energy}
\end{equation}
where $p(\mathbf{o}|\mathbf{z})$ is the underlying directed generative model (or likelihood) that an NGC circuit embodies and $p(\mathbf{z})$ is the prior over its latent variables (or neural activities). $q(\mathbf{s})$ is an approximate posterior distribution over the latent variables $\mathbf{z}$ (given an observation), the choice of which often depends on a set of parameters that are to be optimized. Observe that VFE is a functional that balances two terms---one that encourages improving the likelihood mapping/generative component (or accuracy) while the other term penalizes model complexity (aligning with the idea of Occam's razor) and encourages the underlying model to ensure that the (Kullback-Leibler) divergence between its proxy posterior (or recognition model) and its prior is as small as possible.

Given that we have committed ourselves to viewing our cognitive architecture's foundational neural circuitry as a generative process in of itself, we next briefly specify the form of the generative model that each circuit embodies. A neural circuit's marginal probability has the dependencies:
\begin{align}
    p(\mathbf{z}^0, \dots, \mathbf{z}^L) = p(\mathbf{z}^L) \prod_{\ell=0}^{L-1} p(\mathbf{z}^\ell \mid \mathbf{z}^{\ell+1})
\end{align}
and we further consider the distribution $p(\mathbf{z}^\ell \mid \mathbf{z}^{\ell+1})$ to be a multivariate Gaussian distribution with a mean given by a transformation $f^\ell$ of the latent variables of the level above. In our formulation, $f^\ell$ is a linear map consisting of activation function $\phi^\ell$ and synaptic matrix $\mathbf{W}^\ell(t)$, i.e., {$\mathbf{\bar{z}}^\ell = f^\ell(\mathbf{z}^{\ell+1}) = \mathbf{W}^\ell(t) \cdot \phi^\ell(\mathbf{z}^{\ell+1})$}, where $\cdot$ indicates matrix-vector multiplication. As a result, we arrive at:
\begin{align}
    p(\mathbf{z}^L) = \mathcal N(\mathbf{\bar{z}}^L, \mathbf{\Sigma}^L), \;  p(\mathbf{z}^\ell \mid \mathbf{z}^{\ell+1}) = \mathcal N(f^\ell(\mathbf{z}^{\ell+1}), \Sigma^\ell).
\end{align}
To optimize the generative structure above, all that remains is to specify its energy functional. Due to the particular assumptions made above, coupled with a mean-field approximation that enforces independence among the neurons that compose the circuit (see \cite{friston2010free,ororbia2019lifelong,ororbia2022neural,ororbia2020largescale,salvatori2023brain} for details on derivations/assumptions), we obtain a form of the VFE of Equation \ref{eqn:free_energy} as follows:
\begin{align}
\mathcal{F}(\Theta) = \sum^L_{\ell=0} \frac{1}{2\Sigma^\ell} \sum^{\mathcal{J}_\ell}_{i=1} \big( \mathbf{z}^\ell_i(t) -  \mathbf{\bar{z}}^\ell_i \big)^2\label{eqn:pc_vfe}
\end{align}
where $\mathcal{J}_\ell$ denotes the number of neuronal cells in layer $\ell$ and $\Theta = \{\mathbf{W}^\ell(t),\mathbf{E}^\ell(t)\}^L_{\ell=1}$ (i.e., $\Theta$ is the construct that houses all of the time-evolving generative $\mathbf{W}^\ell(t)$ and message passing $\mathbf{E}^\ell(t)$ synaptic matrices). The above VFE functional highlights two neurobiological commitments made by NGC: 1) there exists a neuronal unit type that specializes in calculating mismatch signals, i.e., the precision-weighted \emph{error neuron} ${\mathbf{e}^\ell(t) = \frac{1}{\Sigma^\ell} \big( \mathbf{z}^\ell(t) -  \mathbf{\bar{z}}^\ell} \big)$\footnote{Note that in this case, we have assumed a fixed scalar precision $\frac{1}{\Sigma^\ell}$ but, in general, this can be constructed to be a learnable synaptic matrix \cite{ororbia2022neural,salvatori2023brain}.}, and 2) a neuron $i$ in one layer (inside of  vector $\mathbf{z}^{\ell+1}$) guesses the activities of another neuron $j$ in layer $\ell$. The dynamics of any neuronal layer, i.e., the \emph{state neurons}, within an NGC circuit concretely follow:
\begin{align}
    \tau_m\frac{\partial \mathbf{z}^\ell(t)}{\partial t} = -\gamma \mathbf{z}^\ell(t) + \mathbf{d}^\ell \odot f_D(\mathbf{z}^\ell(t))  -\mathbf{e}^\ell(t)  \label{eqn:state_update}
\end{align}
where $\odot$ indicates the Hadamard product, $\mathbf{d}^\ell = \mathbf{E}^\ell(t) \cdot \mathbf{e}^{\ell-1}(t)$, i.e., the perturbations produced by error messages that are passed back along feedback synapses $\mathbf{E}^\ell(t)$, $f_D()$ is a neural activity dampening function, and $\tau_m$ is the cellular membrane time constant. The generative synapses within the circuit are adjusted according to the following set of dynamics:
\begin{align}
    \tau_w \frac{\partial \mathbf{W}^\ell(t)}{\partial t} &= -\gamma_w \mathbf{W}^\ell(t) + \mathbf{e}^{\ell-1}(t) \cdot \big(\mathbf{z}^\ell(t)\big)^T \label{eqn:gen_synaptic_update} \\
    \tau_e \frac{\partial \mathbf{E}^\ell(t)}{\partial t} &= -\gamma_e \mathbf{E}^\ell(t) + \mathbf{z}^\ell(t) \cdot \big(\mathbf{e}^{\ell-1}(t)\big)^T\label{eqn:err_synaptic_update}
\end{align}
where $\tau_w$ and $\tau_e$ are generative and error feedback synaptic plasticity time constants, respectively, and $\gamma_w$ and $\gamma_e$ are generative and error feedback synaptic plasticity decay constants, respectively. 
These differential equations indicate that synaptic plasticity is a function of a local update (i.e., a simple two-factor Hebbian rule based on error neuron activity and state unit activity values). See Figure~\ref{fig:ngc} for an illustration of an NGC neural circuit. 

While there are many ways to utilize NGC circuitry, upon presentation of input stimulus $\mathbf{o}(t)$ and possibly top-down context or input vector $\mathbf{c}(t)$ (i.e., meaning we clamp $\mathbf{z}^0(t) = \mathbf{o}(t)$ and $\mathbf{z}^L(t) = \mathbf{c}(t)$) our cognitive system iteratively applies Equation \ref{eqn:state_update} for $T$ steps and then applies Equations \ref{eqn:gen_synaptic_update} and \ref{eqn:err_synaptic_update}, implying that synapses evolve at a slower time-scale than neural activities.

\section{The CogNGen Architecture}
\label{sec:cogngen}

\begin{figure}[!t]
\begin{center}
\includegraphics[width=0.65\linewidth]{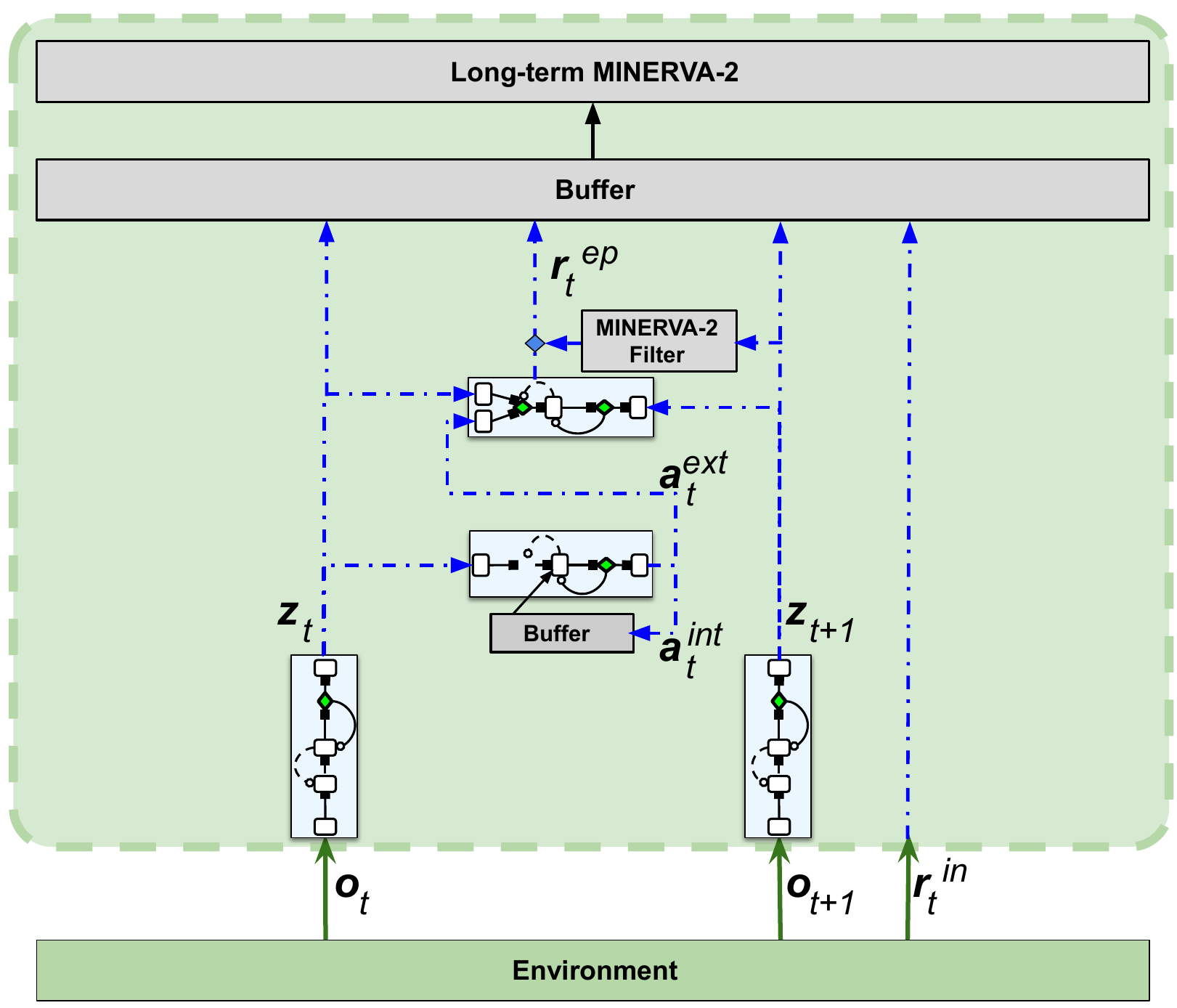}
\end{center}
\caption{The CogNGen control architecture, composed of vector-symbolic memory and free energy minimizing neural circuitry.
}
\label{fig:cogngen_arch}
\vspace{-0.4cm}
\end{figure}

CogNGen's design entails implementing, using either (assemblies of) vector-symbolic memory or free-energy minimizing NGC circuits, a perception system (for one or more modalities), a motor-action module, a procedural memory system, working memory buffers, and a declarative memory. In Figure \ref{fig:cogngen_arch}, we depict a minimal implementation of the CogNGen architecture (i.e., referred to as a ``kernel'', a term we borrow from operating systems design). Note that the diagram depicts CogNGen in ``processing mode'' (i.e., no synaptic update is performed in this mode, only circuit latent states are updated, actions are taken, and internal reward signals are computed/filtered). However, relevant information is stored in a working memory buffer, which interfaces with the long-term MINERVA~2 memory. Transitions are sampled from MINERVA~2 when the ``learning process'' is triggered (i.e., after state $\mathbf{z}_{t+1}$ has been encountered by the perceptual module and stored, samples are replayed from memory to update the motor and procedural circuits). 

\noindent
\textbf{Perception.} In general, an NGC circuit can be constructed to serve the role of the encoder $f_e$ for CogNGen. Doing so would yield the additional advantage that a top-down directed generative model, or decoder $f_g \colon \mathbf{z}_t \mapsto \mathbf{o}_t$, would be learned jointly along with $f_e \colon \mathbf{o}_t \mapsto \mathbf{z}_t$. This is motivated by prior work \cite{ororbia2022neural} showing that NGC learns a good density estimator of data (from which new samples can be ``fantasized''; \cite{ororbia2020continual,ororbia2022neural}). An NGC encoder would, by default, be unsupervised, especially if it is being pre-trained before a task simulation, i.e., the NGC $f_e$ would be trained to predict $\mathbf{o}_t$ given $\mathbf{z}_t$, where $\mathbf{z}_t$ is iteratively crafted by the NGC settling process, and a feedforward model can be trained to amortize the settling process to further reduce computational complexity. Having a decoder  allows for visual interpretation of the distributed representations acquired by CogNGen, since a latent vector $\mathbf{\hat{z}}_t$ (such one as produced by the procedural dynamics model) could be run through the underlying top-down directed generative $f_g$ to produce its corresponding instantiation $\mathbf{\hat{o}}_t$ in observation space.

Another advantage of the above formulation is that, if a task-specific encoder $f_e$ for a modality is available (as in \cite{ororbia2022cogngen}), it may be used alongside or in place of a full NGC encoder circuit. This can simplify and speed up the simulations involving CogNGen, especially if learning a joint perceptual-memory-control system is not the goal, allowing the experimenter to leverage a reliable, stable state representation to design or experiment with various configurations of the CogNGen kernel's other internal sub-systems and observe their impact on the task at hand. 

\noindent
\textbf{Working Memory.} The \textit{self-recurrent slot buffer} (see Figures \ref{FigCMC} \& \ref{fig:cogngen_arch}) serves as the glue that joins the modules of CogNGen together. Working memory in the CMC can be implemented in a variety of ways \cite{laird2017standard}. In ACT-R  \cite{Ritter2019actr}, for example, the mind/brain is understood as consisting of modules connected by buffers, each storing data in a small, finite number of slots. Collectively, the buffers serve as ACT-R's working memory and recurrent slot buffers in CogNGen serve the same purpose, and are inspired by \cite{kruijne2021flexible}'s memory model. The model stores a finite quantity $M_w$ of projected latent state vectors into a set of self-recurrent memory vector slots. Each memory slot in the buffer is represented by $\mathbf{m}^i \in \mathcal{R}^{M_d \times 1}$ ($M_d$ is the dimensionality of the memory slot). Concretely, the self-recurrent slot buffer operates according to the following:
\begin{align}
    \mathbf{k}^i_t &= \mathbf{Q}^i \cdot \mathbf{z}_t, \forall i = 1,...,M_w  \label{eqn:key} \\
    s^i = \mathbf{s}^i &= \frac{1}{|\mathbf{m}^i|} \bigg( \sum_j \lfloor \mathbf{m}^i - \mathbf{k}^i_t \rfloor_{j,1} + \lfloor \mathbf{k}^i_t - \mathbf{m}^i \rfloor_{j,1} \bigg)\label{eqn:match} \\
    \mathbf{m}_t &= \Big[ [\mathbf{m}^1,\mathbf{s}^1],...,[\mathbf{m}^i,\mathbf{s}^i],...,[\mathbf{m}^{M_w},\mathbf{s}^{M_w}] \Big] \label{eqn:value}
\end{align}
where $\mathbf{Q}^i \in \mathcal{R}^{M_d \times D_z}$ is the $i$th random projection matrix (sampled from a centered Gaussian distribution), i.e., there is one projection matrix per working memory slot. The match score for any slot $i$ is $\mathbf{s}^i = \mathcal{R}^{1\times1}$ (a $1\times1$ vector) and thus a scalar $s^i$. The working memory buffers, in effect, calculate the match score between the $i$th key and $i$th slot/value, and then return the entire concatenated contents $\mathbf{m}_t$ of working memory (including match scores). In general, other NGC circuits manipulate these buffers, deciding whether or not to store information within them at any time.

\noindent
\textbf{On Motor-Control.} 
To allow CogNGen to manipulate its world, as well as facilitate internal control (such as modifying a slot buffer), CogNGen is endowed with circuits to drive its actuators. Building on notions of active inference \cite{botvinick2012planning}, we leverage variations of NGC circuits called ``active NGC'' circuits \cite{ororbia2021adapting}. 
Specifically, we design a motor-action model $f_a \colon \mathbf{z}_t \mapsto (\mathbf{c}^{int}_t, \mathbf{c}^{ext}_t)$ (which, in effect, offers some of the functionality provided by the motor cortex) that outputs two control signals at each time step, i.e., internal control signal $\mathbf{c}^{int}_t \in \mathcal{R}^{A_int \times 1}$ and external control signal $\mathbf{c}^{ext}_t \in \mathcal{R}^{A_ext \times 1}$. 
A discrete internal action $a^{int}_t \in \{1,2,...,A_{int}\}$ is read as  $a^{int}_t = \arg\max_{i} \mathbf{c}^{int}_t$ while an external action $a^{ext}_t \in \{1,2,...,A_{ext}\}$ is read as $a^{ext}_t = \arg\max_{j} \mathbf{c}^{ext}_t$ ($A_{int}$ is the number of internal actions and $A_{ext}$ is the number of external actions). 
Action $a^{ext}_t$ impacts the environment that CogNGen is interacting with while action $a^{int}_t$ manipulates coupled memory. At a high-level, any active NGC circuit is trained using a reward signal that is a function of an instrumental term (e.g., sparse problem reward, prior preference circuit), and an epistemic foraging term---a dopamine-like signal produced by procedural memory---to encourage actions that support intelligent exploration.

\noindent
\textbf{Procedural Memory.}
Neuro-behavioral studies find that reward signals are used (by the brain) to evaluate whether or not an action (motor activity) is desirable/undesirable \cite{rangel2008framework}. Action selection is driven by changes in the neural activity of the basal ganglia which estimate the value of the expected reward \cite{hikosaka2006basal}. Motivated by the finding of expected value estimation in the brain, CogNGen's procedural module implements a neural circuit that produces intrinsic (epistemic) signals. At a high level, this machinery facilitates some of the functionality offered by the basal ganglia and procedural memory. 
Specifically, we implement an NGC dynamics model from which a reward signal is calculated as a normalized form of its VFE at time $t$. We couple the dynamics model with a short-term memory module, based on MINERVA~2, 
which adjusts the reward value produced by the dynamics circuit by determining if the currently observed state is familiar or unfamiliar.

\begin{wrapfigure}{r}{0.425\textwidth}
\vspace{-0.455cm}
  \begin{center}
    \includegraphics[width=0.35\textwidth]{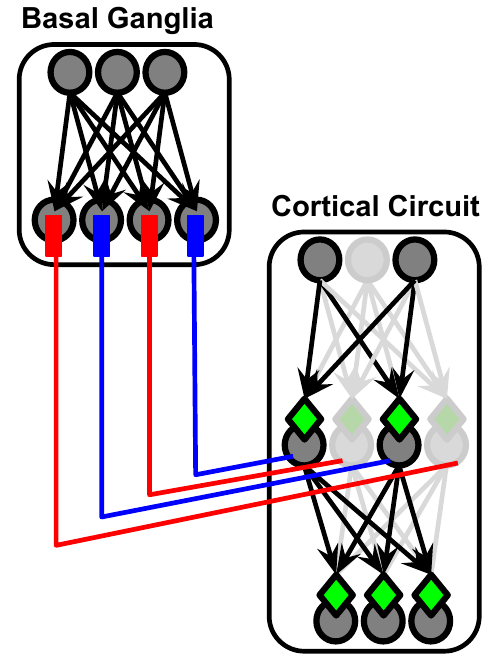}
  \end{center}
  \caption{An element of CogNGen's procedural memory includes circuits that implement the basal ganglia for information routing. Depicted is such a sub-circuit producing gating signals that turn off/on neurons in an NGC perceptual (cortical) circuit.
  }
  \label{fig:basal_ganglia}
  \vspace{-0.5cm}
\end{wrapfigure}

Another important aspect of CogNGen's procedural memory includes sub-circuits that generate ``gating variables'' that dynamically turn on (trigger) or turn off (suppress) sub-networks, i.e., subsets of connected neuronal units within particular neural structures, effectively emulating the basal ganglia's role as a conditional information routing construct \cite{stocco2010conditional,stocco2018biologically}. These sub-circuit structures are learned through various forms of competitive Hebbian learning (which is particular design choice influenced by its ability to prevent catastrophic forgetting from happening within the basal ganglia itself; \cite{ororbia2019lifelong}). While this element of CogNGen's procedural memory has not yet been evaluated in the context of the whole system itself, prior effort has shown that complementary basal ganglia-perceptual systems work well in the face of online task-free continual learning \cite{ororbia2021continual,ororbia2019lifelong}. 
We remark that, while CogNGen focuses on providing task-switching level executive functionality through the basal ganglia/procedural memory, other candidate brain structures that could provide this dynamic neural region suppression/gating include the prefrontal cortex \cite{rougier2002learning}.

\noindent
\textbf{Long-Term Memory.} For CogNGen, we adopt \cite{Collins2020minerva}'s approach and model both working and declarative memory using MINERVA~2. Our working  MINERVA~2, like the slot buffer, is cleared after a task is completed (e.g., a maze is solved), whereas the contents of the declarative MINERVA~2 (which serves as the episodic memory in CogNGen) persist.

Much as is done in (deep) reinforcement learning (RL), in order to improve the stability and convergence of networks trained over many episodes \cite{mnih2015human}, CogNGEN leverages its MINERVA~2 episodic memory by sampling from it, inducing an approximate form of experience replay. This is motivated by early studies of rats where neural replay sequences were detected in the hippocampus \cite{skaggs1996replay} during rest---it was found that ``place'' cells spontaneously/rapidly fired so as to represent the previous paths traversed by the rats when they had been awake. These ``replay'' sequences lasted only a fraction of a second but essentially covered several seconds of real-world experience. Similar replay effects have been been detected in human subjects \cite{kurth2016fast}, providing further neurobiological justification of the replay buffer used in modern-day RL systems. 
In CogNGen, information is transferred into this MINERVA~2 memory through an intermediate working buffer (which contains partial experience information encountered over time, such as state, action taken, and reward). At various points in time, when CogNGen is not adjusting synapses online, memories are replayed sequentially from the episodic memory and the confabulated patterns are used to induce an extra update to parameters. 


\subsection{Simulation Results}
\label{sec:simulations}

\begin{table*}[t!]
\begin{center}
\begin{tabular}{c c c c c}
    \includegraphics[height=0.75in]{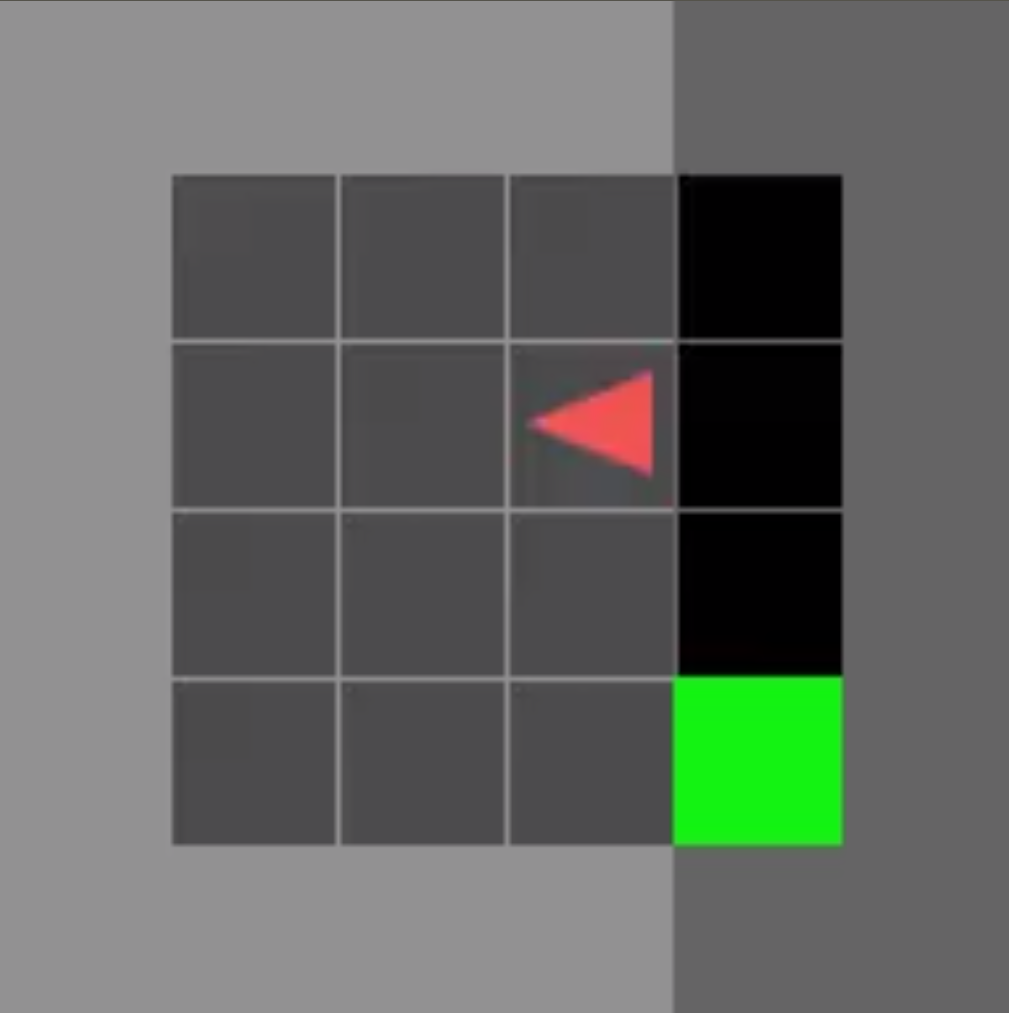} &
    \includegraphics[height=0.75in]{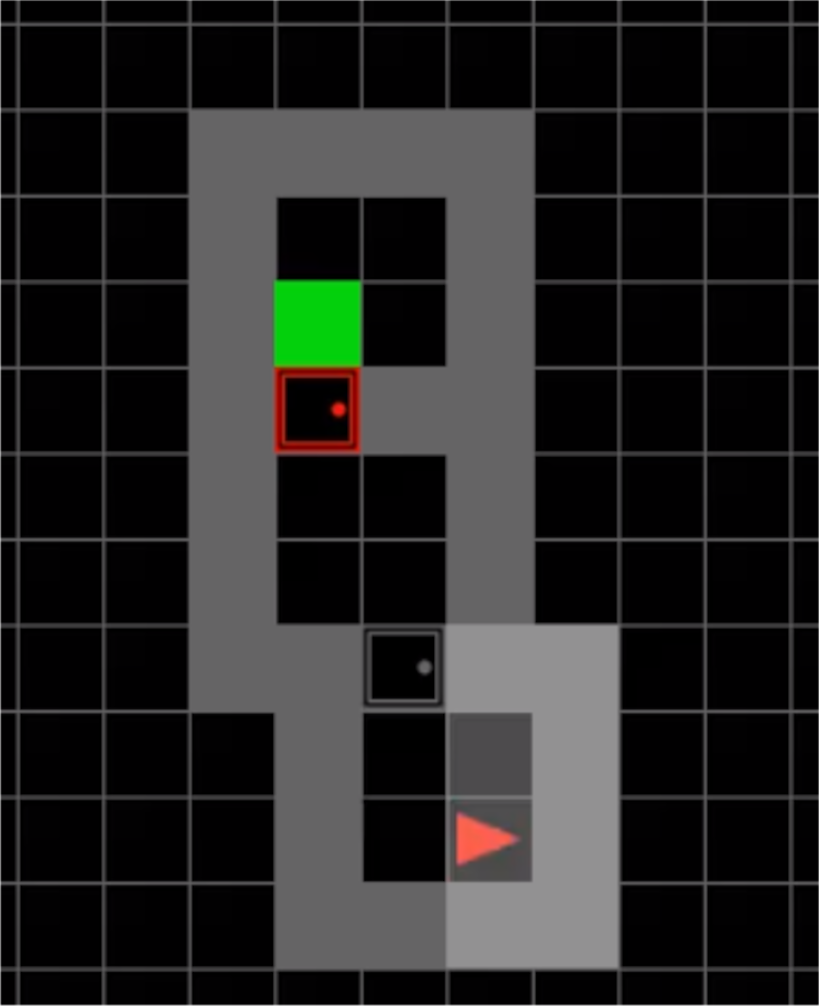} &    \includegraphics[height=0.75in]{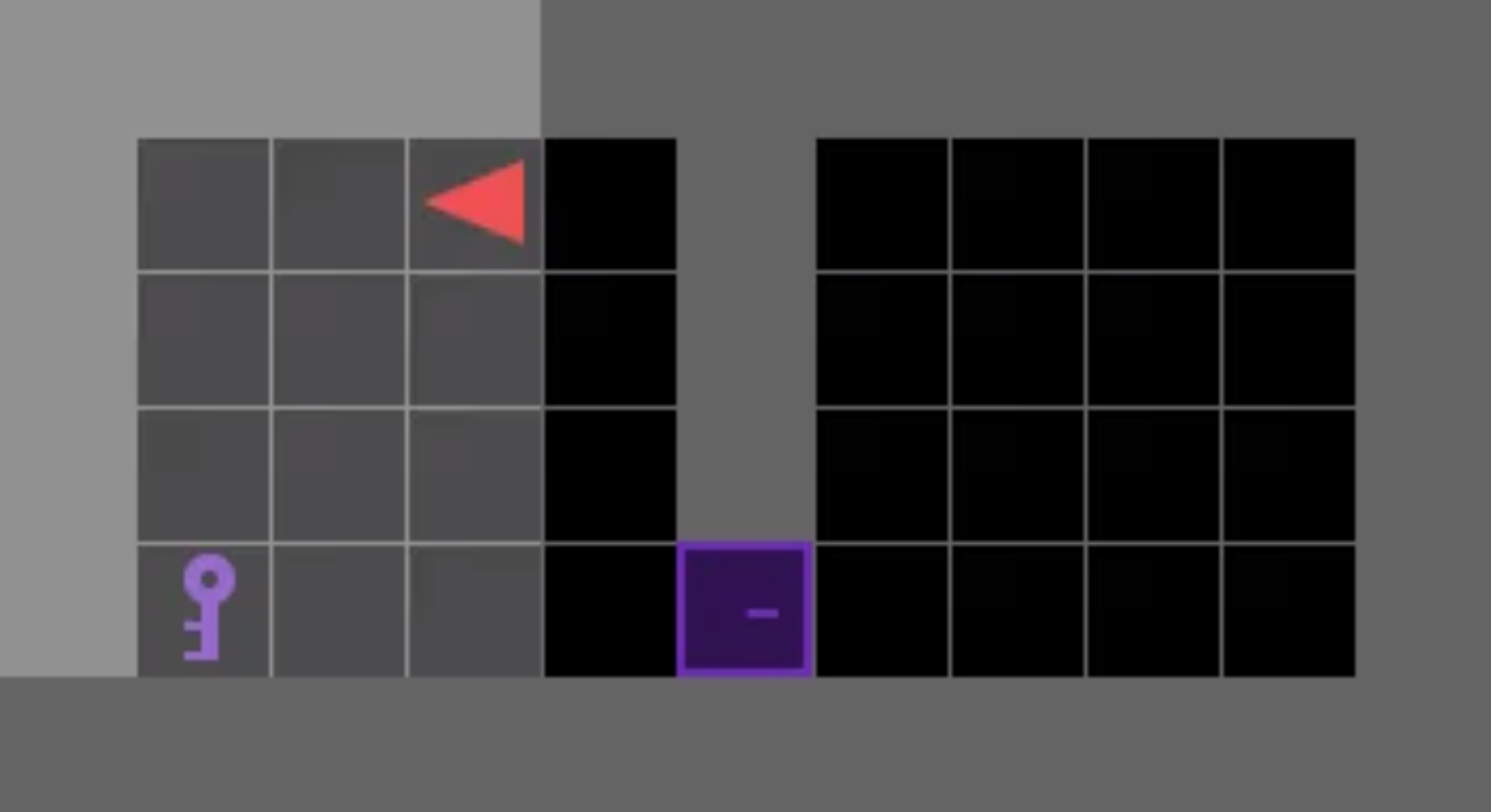} &
    \includegraphics[height=0.75in]{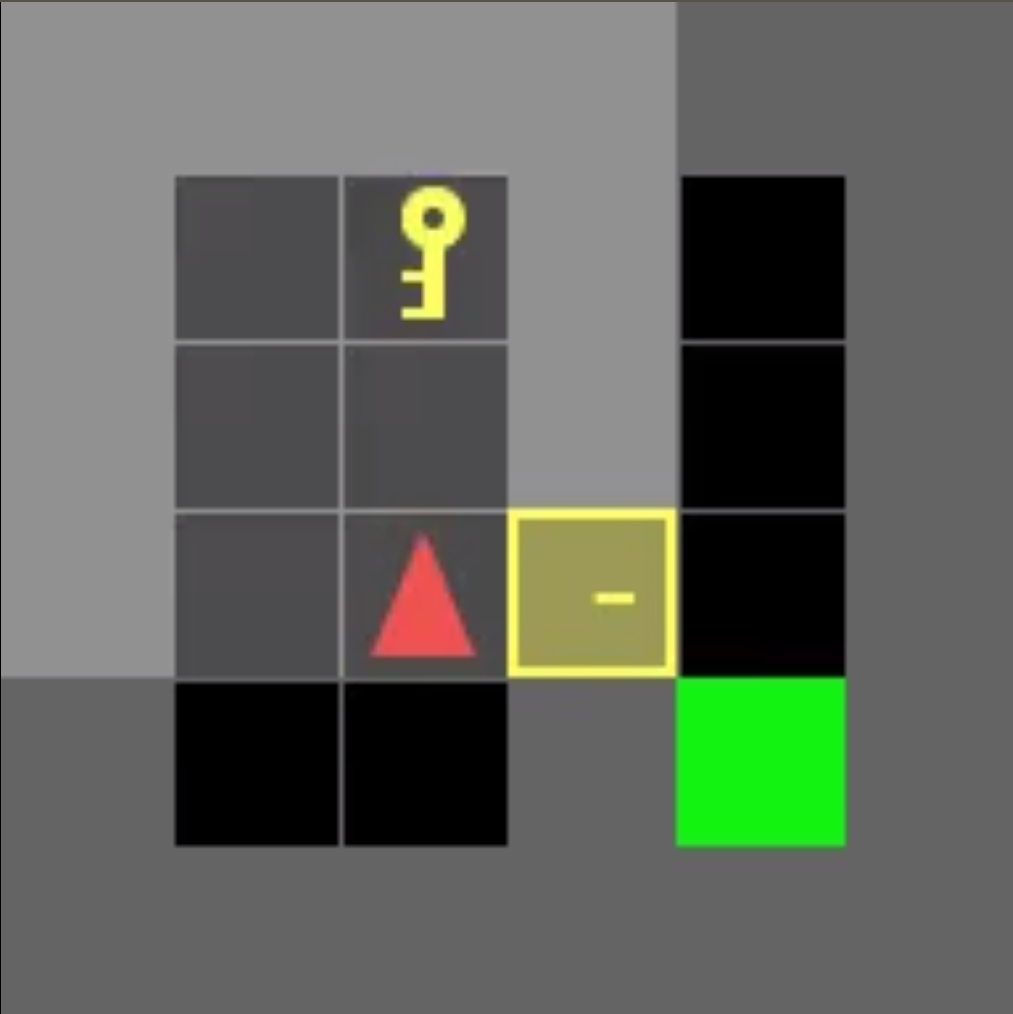} &
    \includegraphics[height=0.75in]{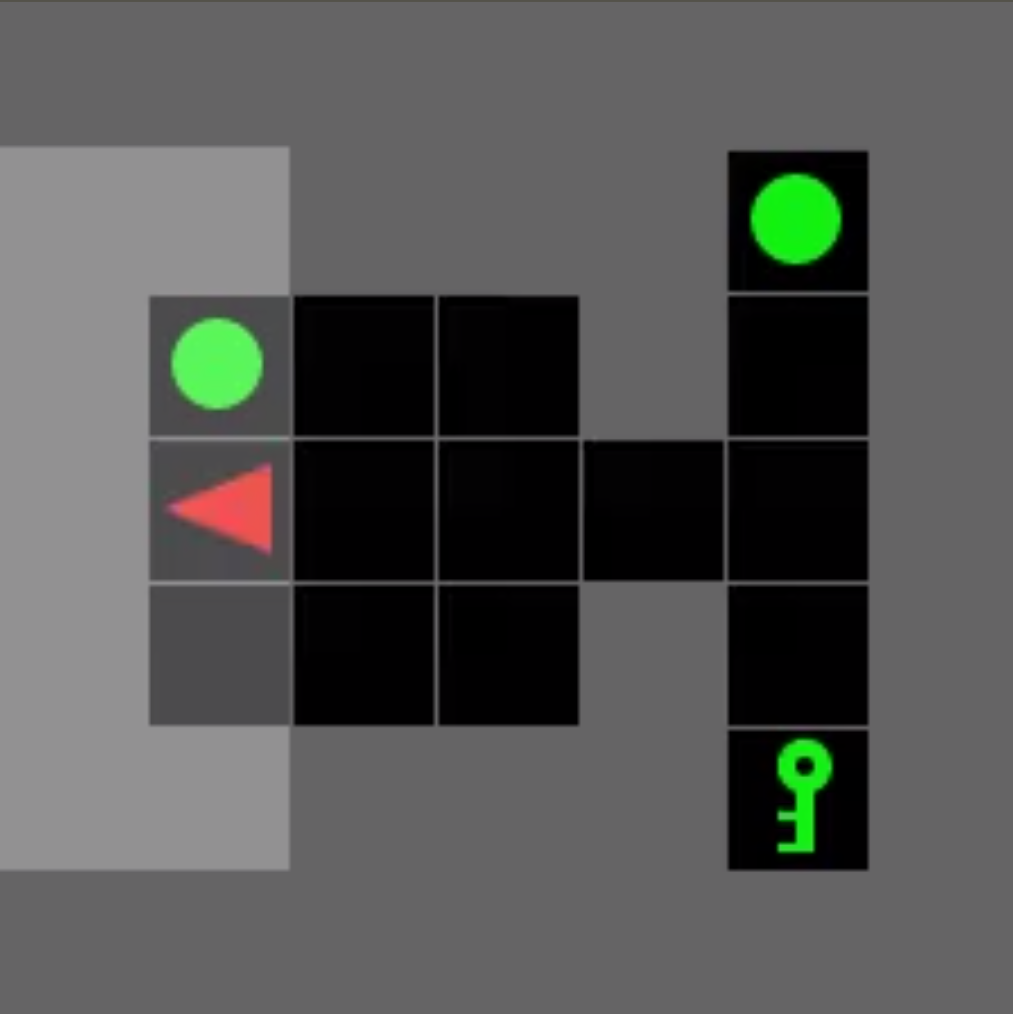} \\
\end{tabular}
\begin{tabular}{l | c c c c c | c c c c c} 
     \hline
      & \multicolumn{5}{|c}{Average Success Rate} & \multicolumn{5}{|c}{Average Episode Length} \\
             & R6x6 & MR & Unl & DK & Mem & R6x6 & MR & Unl & DK & Mem \\ 
     \hline\hline
     DQN & $99.50$ & $0.00$ & $0.00$ & $0.00$ & $40.0$ & $9.31$ & $100.0$ & $100.0$ & $100.0$ & $41.14$\\ 
     \hline
     RnD & $\textbf{100.00}$ & $90.00$ & $\textbf{100.0 }$& $\textbf{100.00}$ & $48.5$ & $3.50$ & $31.46$ & $4.08$ & $3.71$ & $2.78$\\ 
     \hline
     BeBold DQN-CNT & $\textbf{100.00}$ & $98.00$ & $\textbf{100.0}$ & $\textbf{100.00}$ & $48.0$ & $3.98$ & $23.51$ & $4.46$ & $3.93$ &$2.92$\\
     \hline
     CogNGen & $\textbf{100.00}$ & $\textbf{98.50}$ & $\textbf{100.0}$ & $\textbf{100.00}$ & $\textbf{98.5}$ & $3.90$ & $23.41$ & $4.15$ &  $5.48$ & $2.96$\\ 
     \hline
\end{tabular}
\end{center}
\caption{\textit{Above:} Visualization of Gym Mini-Grid tasks that CogNGen has been empirically demonstrated on. From left to right: random room escape (R6x6), procedural multi-room task (MR), door-unlock (Unl), door-key (DK), and memory task (Mem). 
\textit{Below:} We present results for the average success rate (\%) over the last $100$ episodes and average episode length (\% of maximum/worst-case episode length - closer to $0$ is better/more efficient) over the last $100$ episodes.}
\label{table:cogngen_tasks}
\vspace{-0.4cm}
\end{table*}

Empirically, CogNGen has already been demonstrated to perform well \cite{ororbia2022cogngen,ororbia2023maze} on a set of tasks taken from the grid-world environment benchmark known as Mini-GridWorld (an OpenAI Gym extension; \cite{gym_minigrid}). Taken together, these particular tasks, with results depicted in Table \ref{table:cogngen_tasks}, provide a useful suite of controlled simulations of varying complexity that would test the viability and effectiveness of initial prototypes of the CogNGen system. We have compared the CogNGen to several baseline models: a standard deep Q-network (DQN; \cite{mnih2015human}), a DQN that equipped with random network distillation (RnD; \cite{burda2018exploration}), and a DQN that learns through the BeBold exploration framework (BeBold DQN-CNT; \cite{zhang2020bebold}). See \cite{ororbia2022cogngen,ororbia2023maze} for simulation environment setup and additional experimental details. 

\section{On the Value of a Neuro-mimetic CMC} 
\label{sec:value_of_neurocmc}

Although the current form of CogNGen is not without its limitations, there are already many advantages to the cognitive control architecture developed thus far. Notably, when taking in the totality of its vector-symbolic memory and predictive coding elements, CogGNen can be written down as optimizing, in a modular fashion, a single free energy function; this means that the entire system is minimizing a sum of local functionals, including the vector-symbolic memory components (in terms of their Hopfield energies) and the NGC circuits (in terms of their VFEs). Formally, this global VFE is: $\mathcal{F}(\Theta) = \sum_j \mathcal{F}_j(\Theta_j)$, where $\Theta_j$ contains the plastic synapses associated with structure $j$ (e.g., this could be a memory module within declarative memory or a NGC circuit portion of the motor cortex) and $\mathcal{F}_j$ is its  associated (free) energy functional.

Desirably, at the heart of CogNGen is the story of online learning and adaptation, allowing one to characterize its global decisions (with respect to tasks) from not only the behavioral level but also the neural dynamics level. In addition, it is the particular form of CogNGen's neural dynamics, which follow the gradient flow of a variational free energy, that not only connect well to Bayesian brain theory (and the grander picture of predictive processing as not only a mechanistic description of neural activity but as a general principle underlying brain structure and function) but also permit flexible design choices such as the encouragement of strong sparsity in the neural activities. In addition, CogNGen is a small piece of evidence that a heterogeneous mixture of learning paradigms---Hebbian adaptation in the memory modules, local error-driven plasticity in the neural generative coding (NGC) circuits, and competitive Hebbian learning for the basal ganglia construct---can yield a coherent platform for crafting brain-motivated structures to instantiate the various core modules of the Common Model of Cognition. Beyond this, the neural dynamics of NGC generalize well to the level of spiking neuronal cells \cite{ballard2000single,ororbia2019spiking}, offering a possible transition between CogNGen's current story of adaptation through coarse-grained rate coded dynamics to finer-grained spiking electro-physiological dynamics.

\subsection{Future Work}
\label{sec:future_work}

CogNGen is a functional implementation of the Common Model of Cognition and is able to learn a sample of reinforcement learning tasks at a rate comparable to or exceeding competing deep Q-learning networks (see \cite{ororbia2022cogngen,ororbia2023maze}). However, CogNGen, as implemented, is only a prototype. Our goal is to develop CogNGen to be a general purpose framework able to learn a wider variety of tasks and to learn different tasks consecutively without catastrophic forgetting. Learning tasks with continuous-valued perceptual-motor environments may require CogNGen to incorporate more complex  perceptual and motor modules, for which we would use additional NGC circuits.

The main obstacle in scaling CogNGen to larger tasks, and consecutive sequences of distinct tasks, is memory architecture. In our current work, we are refining CogNGen's memory. The first step is a switch from using the MINERVA~2 memory retrieval (Equation~\ref{eqn:minerva_retrieve}) to the more efficient softmax retrieval (Equation~\ref{eqn:transformer_retrieve}) of modern Hopfield networks and transformers). The second step is to allow CogNGen to learn to how to control and modulate the contents of the short-term and long-term MINERVA~2 models / Hopfield networks in the same manner that CogNGen learns to control the contents of the slot buffers. At present, the contents of the MINERVA~2 models are hardwired to be histories of recent observations. Allowing CogNGen to control memory would allow it to pay more attention to some observations than others as well as learn representations for memories.

If in scaling CogNGen to larger tasks or more diverse tasks, we need more long-term memory storage than what is provided by modern Hopfield networks / MINERVA~2, we may explore the episodic / semantic memory distinction and add a separate semantic memory, a model that scales with the number of concepts or categories rather than the number of new episodes (e.g., \cite{kanerva1988sparse,Kelly2020hdm}).

\section{Conclusion}
\label{sec:conclusion}

CogNGen is a proof-of-concept that the Common Model of Cognition provides an appropriate blueprint for designing a modern, deep neural cognitive architecture capable of learning to complete arbitrary tasks. CogNGen commits to using free energy minimization and Hebbian learning rather than the standard back-propagation of errors (backprop), a choice motivated by:  (1) the lack of neuro-scientific evidence for backprop; (2) the recent success of Hebbian-like neuro-circuitry in the attention heads of transformer models; (3) the long history of success of Hebbian learning in computational cognitive modelling; and (4) the aim of designing an architecture that can more easily avoid catastrophic forgetting, allowing it to, in principle, be a more general learner (as is desirable for modelling learning in human and animals). We are excited to push forward this architecture with further refinements and more simulations, and equally excited for other Common Model-inspired approaches to designing intelligent agents using deep learning.

\subsection*{Acknowledgments}
We acknowledge the support of the Natural Sciences and Engineering Research Council of Canada (NSERC), [DGECR-2023-00200] and Cisco Research Gift Award \#26224.

\bibliographystyle{acm}
\bibliography{Kelly_refs,Ororbia_refs}

\end{document}